\documentclass[a4paper]{aa}

\usepackage{graphicx,float}
\usepackage{latexsym,amsmath,amssymb}
\usepackage{natbib}
\bibpunct{(}{)}{;}{a}{}{,}

\def \xmm {XMM-Newton}
\def \src {EXO\,0748$-$676}
\def \degmark{^\circ}
\def \nh {N${\rm _H}$}
\def\ang {$\rm\AA$}

\def \hcm {\hbox {\ifmmode $ atom cm$^{-2}\else atom cm$^{-2}$\fi}}
\def \arcmin {\hbox{$^\prime$}}
\def \arcsec {\hbox{$^{\prime\prime}$}}

\def \chisq {$\chi ^{2}$}
\def \rchisq {$\chi_{\nu} ^{2}$}
\def\approxgt{\mathrel{\hbox{\rlap{\lower.55ex \hbox {$\sim$}}
        \kern-.3em \raise.4ex \hbox{$>$}}}}
\def\approxlt{\mathrel{\hbox{\rlap{\lower.55ex \hbox {$\sim$}}
        \kern-.3em \raise.4ex \hbox{$<$}}}}

\newcommand{\mc}{\multicolumn}
\newcommand {\Msun}{M_\odot}

\newcommand {\phind} {$\Gamma$}

\def \mxb {MXB\,1658$-$298}
\def \ks {KS\,1731$-$260}
\def \xte {XTE\,J1701$-$462}

\newcommand {\unitplnorm} {ph.~keV$^{-1}$~cm$^{-2}$~s$^{-1}$}

\def \nh {$N{\rm _H}$}

\def \kpl {$k_{\rm pl}$}

\begin{document}

\title{\xmm\ observations of the low-mass X-ray binary \src\ in quiescence}

\author{M. D{\'i}az Trigo\inst{1} \and L. Boirin\inst{2} \and E. Costantini\inst{3} \and M. M\'endez\inst{4} \and A. Parmar\inst{5}
}
\institute{
ESO, Karl-Schwarzschild-Strasse 2, 85748 Garching bei M\"unchen, Germany
\and
       Observatoire Astronomique de Strasbourg, 11 rue de l'Universit\'e,
       F-67000 Strasbourg, France
\and
	SRON, Netherlands Institute for Space Research,
         Sorbonnelaan 2, 3584 CA  Utrecht, The Netherlands
\and
	Kapteyn Astronomical Institute, University of Groningen, Postbus 800, 9700 AV Groningen, The Netherlands
\and
       XMM-Newton Science Operations Centre, Science Operations and Data Systems Division,
       Research and Scientific Support Department of ESA, ESAC, Apartado 50727, 28080
       Madrid, Spain
}

\date{Received ; Accepted:}

\authorrunning{D{\'i}az Trigo et al.}

\titlerunning{XMM-Newton observations of \src\ in quiescence}

\abstract{The neutron star low-mass X-ray binary \src\ started a transition from outburst to quiescence in August 2008, after more than 24~years of continuous accretion. The return of the source to quiescence has been monitored extensively by several X-ray observatories. Here, we report on four \xmm\ observations elapsing a period of more than 19 months and starting in November 2008. The X-ray spectra show a soft thermal component which we fit with a neutron star atmosphere model. Only in the first observation do we find a significant second component above $\sim$3~keV accounting for $\sim$7\% of the total flux, which could indicate residual accretion. The thermal bolometric flux and the temperature of the neutron star crust decrease steadily by 40 and 10\% respectively between the first and the fourth observation. At the time of the last observation, June 2010, we obtain a thermal bolometric luminosity of 5.6\,$\times$\,10$^{33}$ (d/7.1 kpc)$^2$ erg s$^{-1}$  and a temperature of the neutron star crust of 109~eV. The cooling curve is consistent with a relatively hot, medium-mass neutron star, cooling by standard mechanisms. From the spectral fits to a neutron star atmosphere model we infer limits for the mass and the radius of the neutron star. We find that in order to achieve self-consistency for the NS mass between the different methods, the value of the distance is constrained to be
$\approxlt$\,6~kpc. For this value of the distance, the derived mass and radius contours are consistent with a number of EoSs with nucleons and hyperons.

} \maketitle

\section{Introduction}
\label{sect:intro}

\src\ is one of the most interesting transient low mass X-ray binaries (LMXBs)
observed to date.
It was discovered with
EXOSAT \citep{0748:parmar86apj}. It shows 8.3 minute X-ray eclipses
every 3.82 hr, irregular dipping activity and type I X-ray bursts
\citep{0748:parmar86apj, 0748:gottwald86apj}. 

Assuming Roche lobe
overflow, a main-sequence companion, and a 1.4~$\Msun$ neutron star (NS)
\citet{0748:parmar86apj} constrained first the inclination and companion mass to
$75\degmark < i < 83\degmark$ and 0.45~$\Msun$, respectively. 
From measurements of the radial velocity of the companion, using the technique of
Doppler tomography on optical spectra, \citet{0748:munozdarias09mnras} 
derived a mass for the neutron star
between 1 and 2.4~$\Msun$ and a mass ratio 0.11~$<$~$q$~$<$~0.28. 
\citet{0748:bassa09mnras} used the same technique to analyse observations performed after a significant cessation of accretion activity in \src\ was detected in 2008 and placed a lower limit on the NS mass of 1.27~$\Msun$ and constrained the mass ratio to 0.075~$<$~$q$~$<$~0.105. For a 1.4~$\Msun$ NS this translates into a mass for the companion star of 0.11~$<$~$M_2$~$<$~0.15~$\Msun$.

Quasi-periodic oscillations (QPOs) were
discovered during persistent emission at Hz and kHz frequencies
\citep{0748:homan99apjl, 0748:homan00apj}. Burst oscillations were
first detected at 45~Hz in an average of 38 type-I X-ray bursts by \citet{0748:villarreal04apj}, which were interpreted as the spin frequency of
the neutron star. However, \citet{0748:galloway10apjl} reported oscillations in the rising phase of two type-I X-ray bursts at a frequency of 552~Hz and concluded that the latter is the spin frequency of the NS, while the 45~Hz oscillation may arise in the boundary layer between the disc and the NS \citep{0748:balman09atel} or it could be a statistical fluctuation \citep{0748:galloway10apjl}. 

From the analysis of a photospheric radius expansion (PRE) X-ray burst \citet{0748:wolff05apj} estimated the distance of \src\ to be 5.9--7.7~kpc. Extending the analysis to several type-I X-ray bursts, \cite{dist:galloway08apjs} and \citet{dist:galloway08mnras} estimated a distance of 7.4\,$\pm$\,0.9~kpc and 7.1\,$\pm$\,1.2~kpc, respectively, the latter value taking into account the touch-down flux and the high-inclination of \src.

\src\ is the only source for which
gravitationally redshifted absorption lines during X-ray bursts have been reported \citep{0748:cottam02nat}. A
gravitational redshift of z = 0.35 at the star surface was inferred
from these measurements, which translates to a mass-to-radius ratio of
M/R = 0.152 $\Msun$/km \citep{0748:cottam02nat}. {\it This provides an
empirical constraint on the equation of state (EoS) of dense, cold nuclear
matter.} \citet{0748:ozel06nat} used the value
of the gravitational redshift, combined with an estimate of the
stellar radius, to infer a mass for the NS of M\,$\approxgt$\,1.82\,$\Msun$ for a stellar radius of R\,$\approxgt$\,12~km in the coordinate
frame of the NS. \citet{0748:rauch08aa} regarded the line
identifications from \citet{0748:cottam02nat} as highly
uncertain from computations of LTE and non-LTE NS model atmospheres
and derived instead
a redshift of z = 0.24 with an alternative line identification and a 
NS radius of R = 12--15~km for the mass range M = 1.4--1.8\,$\Msun$.
The spectral features seen in
the burst spectra from the initial data \citep{0748:cottam02nat} could
not be reproduced in the 68 burst spectra from a longer observation in 2003
\citep{0748:cottam08apj}. This could be due to changes in the ionisation
conditions in the NS photosphere \citep{0748:cottam08apj}. However, it remains difficult
to reconcile the narrowness of the absorption lines detected in the initial data if they
originate in the surface of the NS after the recent rapid spin reported by  \citet{0748:galloway10apjl} 
\citep[e.g.][]{0748:lin10apj}.

An alternative method to set constraints to the EoS of NSs  consists in deriving their mass and radius 
from X-ray spectral fitting to the thermal emission component after accretion ceases. The theory behind
this method is the so-called deep crustal heating of accreting NSs, according to which the nuclear 
ashes sink in the NS crust under the weight of newly accreting matter and with increasing pressure, undergo 
a sequence of nuclear transformations (pycnonuclear reactions) accompanied by heat deposition \citep[e.g][]{haensel90aa}. The heat gained via this process is lost during quiescent episodes, resulting in thermal emission from the NS surface. The equilibrium temperature is set by the temperature of the NS core, which in turn depends on the long-term time-averaged mass accretion rate of the NS and the extent to which the core is able to cool via neutrino emission \citep{rutledge02apj}. Thus, fitting the thermal component NS atmosphere models allows us to gain insight in the structure and composition of the NS crust and core, {\it if a reliable distance estimate is available}. The significant cease of accretion of \src\ in 2008 (see below) opened the possibility of  constraining its mass and radius using this method. 

\citet{0748:wolff08atel} reported on RXTE observations of \src\ during August 2008 showing the
lowest flux level, 6.8\,$\times$\,10$^{-11}$ ergs~cm$^{-2}$~s$^{-1}$ between 3 and 12~keV, 
since the beginning of the RXTE mission in 1996. The flux level appeared
near the detection threshold of the RXTE All Sky Monitor (ASM) during
September 2008. Subsequent $Swift$ and RXTE observations of the source
in October 2008 confirmed that accretion had largely ceased and \src\
was returning to quiescence \citet{0748:wolff08atelb}. 

Following the significant cease of accretion, we proposed four long (30--100~ks) observations of \src\ with \xmm, three equally spaced during the first year and a fourth one in the second year. With the sensitivity provided by these observations we aimed to obtain new constraints to the EoS of dense matter via 3 complementary methods: analysis of the heating and cooling curves and spectral fitting with NS atmosphere models. The cease of accretion of \src\ has been also monitored with {\it Chandra} and {\it Swift} \citep{0748:degenaar09mnras,0748:degenaar10mnras}. However, the shorter observations and the lower sensitivities of these observatories compared to \xmm\ limit their capability to obtain significant constraints to the mass and radius of \src. In this paper, we report on our monitoring program with \xmm\ (the first of these observations has been analysed also by \citet{0748:zhang10mnras}). The set of four observations analysed here constitute the most sensitive dataset of \src\ after its return to quiescent state. 

\section{Observations}
\label{sec:observations}

The XMM-Newton Observatory \citep{jansen01aa} includes three
1500~cm$^2$ X-ray telescopes each with an EPIC 
(0.1--15~keV) at the focus.  Two of the EPIC imaging
spectrometers use MOS CCDs \citep{turner01aa} and one uses pn CCDs
\citep{struder01aa}. The RGSs \citep[0.35--2.5~keV,][]{denherder01aa} 
are located behind two of the telescopes. In addition, there is a co-aligned 30~cm diameter 
Optical/UV Monitor telescope \citep[OM,][]{mason01aa}, 
providing simultaneously coverage with the X-ray instruments.
Data products were reduced using the Science Analysis
Software (SAS) version 10.0. We present here the analysis 
of EPIC data, RGS data from both gratings and OM data.

\begin{table*}
\begin{center}
\caption[]{\xmm\ observations of \src\ since November 2008. $T$ is
the exposure time and $C$ the EPIC pn 0.3--10~keV source persistent net count rate. In all cases the Full Frame Mode with the Thin filter was used for all the EPIC cameras, except for Obs~0560180701, for which the Medium filter was used.}
\begin{tabular}{cclcccc}
\hline \noalign {\smallskip}
Observation & \mc{1}{c}{Date} & Instrument & $T$  & $C$ \\
ID   &                 &     & (ks) & (s$^{-1}$) & \\
\hline \noalign {\smallskip}
0560180701 & 2008 November 6  & pn  & 29 & $0.648 \pm 0.006$ \\
	   &		       & MOS1 & 29 & $0.169 \pm 0.003$\\
	   &		       & MOS2 & 29 & $0.160 \pm 0.002$  \\
0605560401 & 2009 March 17     & pn   & 41 & $0.525 \pm 0.004$ \\
	   &		       & MOS1 & 43 & $0.105 \pm 0.002$ \\ 
	   &		       & MOS2 & 42 & $0.131 \pm 0.002$  \\
0605560501 & 2009 July 1 & pn   & 100 & $0.473 \pm 0.004$  \\
	   &		       & MOS1 & 101 & $0.117 \pm 0.002$  \\
	   &		       & MOS2 & 101 & $0.115 \pm 0.002$  \\
0651690101 & 2010 June 17 & pn   & 96 & $0.457 \pm 0.004$  \\
	   &		       & MOS1 & 98 & $0.110 \pm 0.001$ \\
	   &		       & MOS2 & 98 & $0.106 \pm 0.001$  \\
\noalign {\smallskip} \hline \label{tab:obslog}
\end{tabular}
\end{center}
\end{table*}

Table~\ref{tab:obslog} is a summary of the \xmm\ observations reported in this paper. 
 All the observations were performed using the EPIC Full Frame mode. In this
mode all the CCDs are read out and the full field of view of the pn (MOS) is covered
every 73~ms (2.6~s). The source count rate is well below the threshold
of 6 (0.7)~s$^{-1}$ where pile-up effects become important for the pn (MOS) cameras. 
We extracted source events from a circle of 
radius between 37 and 47\arcsec\ centered on the PSF core and background events
from a circle with the same radius centered well away from the
source. Ancillary response files were generated using the SAS task
{\tt arfgen}. Response matrices were generated using the SAS task {\tt rmfgen}.
Background subtracted light curves were generated with the SAS task
{\tt epiclccorr}, which corrects for a number of effects like
vignetting, bad pixels, PSF variation and quantum efficiency and
accounts for time dependent corrections within a exposure, like dead
time and GTIs.

The SAS task {\tt rgsproc} with
the option {\it spectrumbinning}=lambda
was used to produce calibrated RGS event lists, spectra, and response
matrices in an uniform wavelength grid space. 
We also chose the option {\it keepcool}=no to discard single
columns that give signals a few percent below the values expected from
their immediate neighbours. Such signals could be mis-interpreted as
weak absorption features in spectra with high
statistics.
We generated RGS light curves with the SAS task 
{\tt rgslccorr}. 
We used the SAS task {\tt rgscombine} to add spectra
from the same RGS and order of different observations and to add 
spectra from RGS1 and RGS2 for one single observation in order to reduce
the statistical errors.

The OM was operated in Image Mode with UVW1, UVM2 and UVW2 filters in Obs~1 and in Image+Fast Mode with U filter in Obs~2--4. 
In the Image mode the instrument produces images of the entire 17\arcmin\ x
17\arcmin\ FOV with a time resolution between 800 and 5000~s. In the Fast mode, event
lists with a time resolution of 0.5~s from a selected 11\arcsec\ x
11\arcsec\ region are also stored. The SAS tasks {\tt omichain} and {\tt omfchain} were used to extract
images and light curves of \src.

\section{Results}

\subsection{X-ray lightcurves}

\begin{figure*}[!ht]
\centerline{\includegraphics[angle=0]{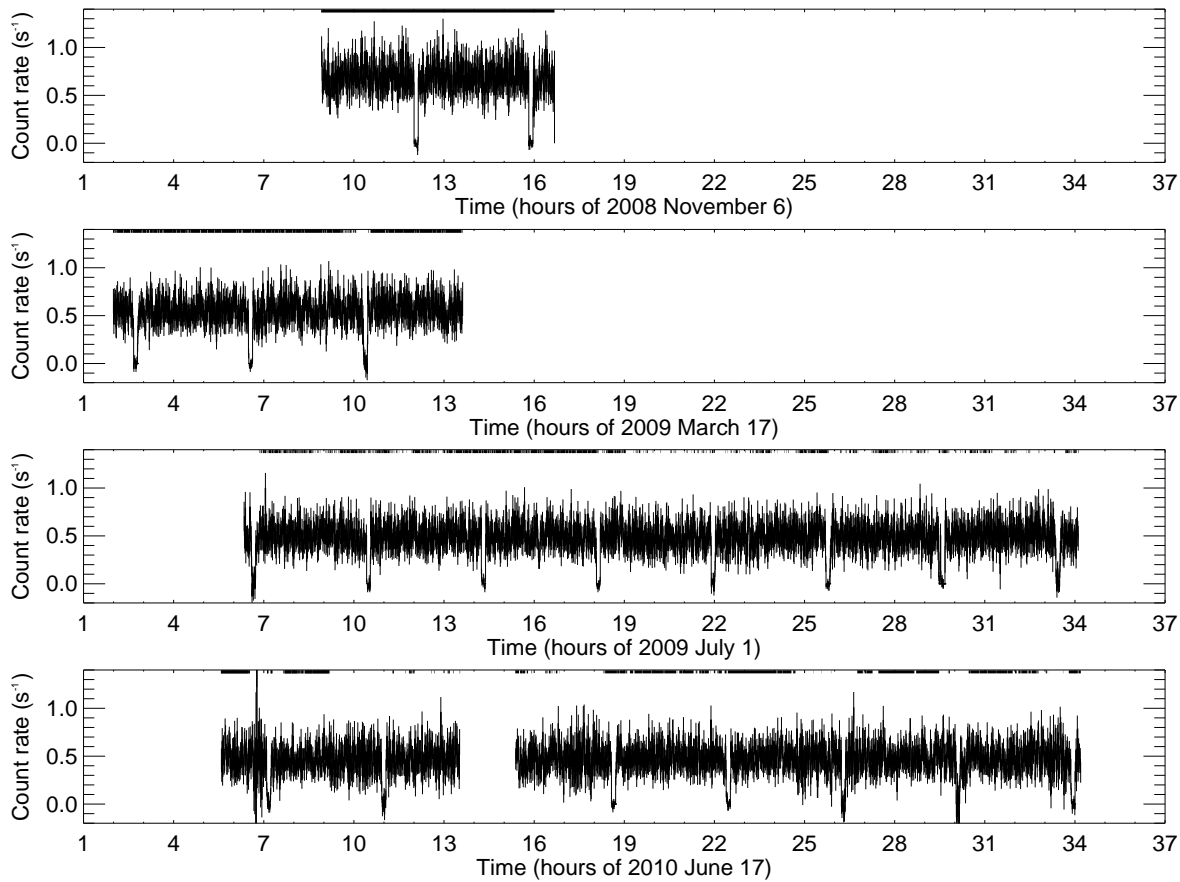}}
\caption{0.3--10~keV EPIC pn background subtracted lightcurves with a binning of 50~s.
The narrow eclipses are clearly evident. The thick horizontal lines mark the good time intervals used for spectral extraction.
\label{fig:lightcurves}}
\end{figure*}

\begin{figure*}[!ht]
\centerline{\includegraphics[angle=0]{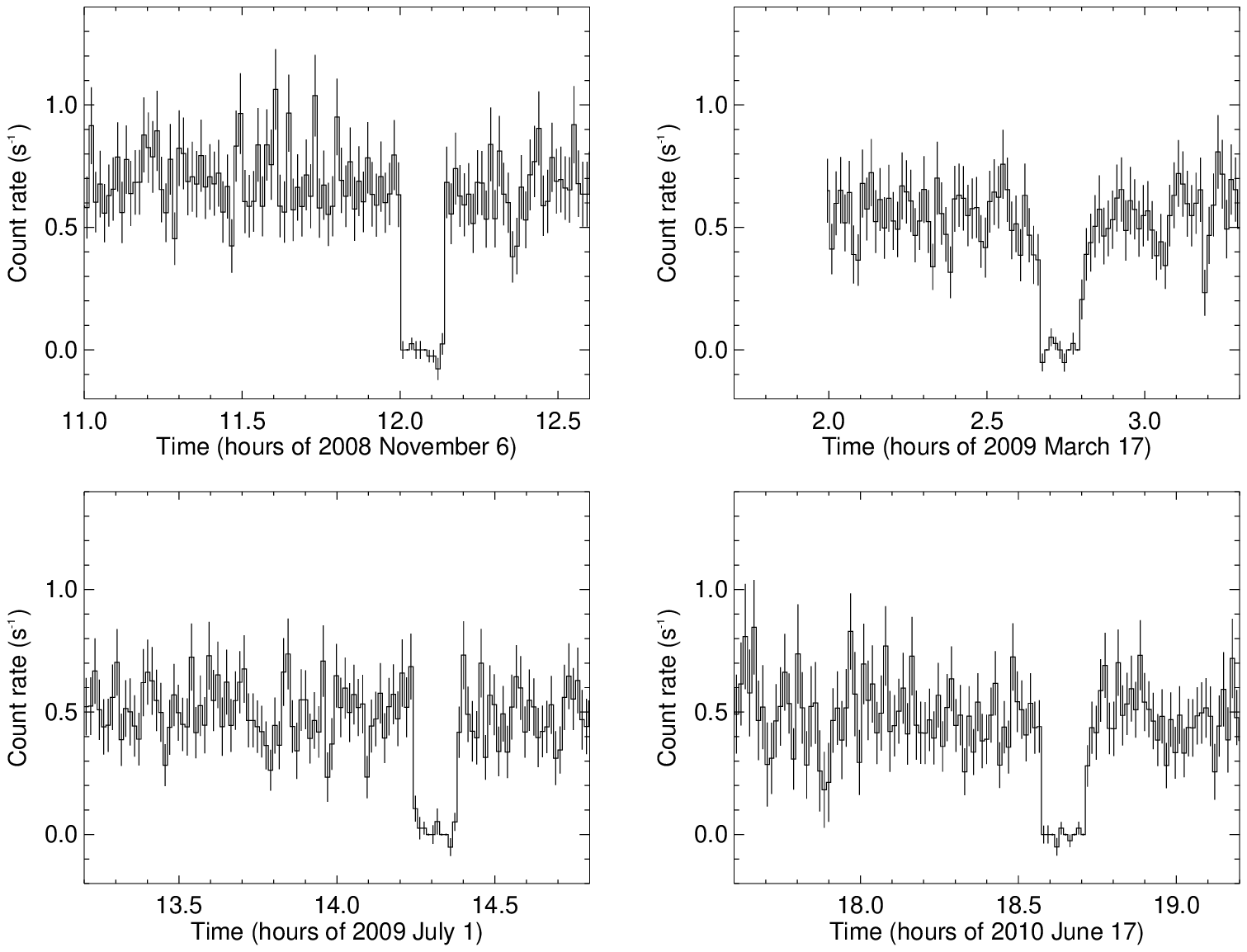}}
\caption{Detail of 0.3--10~keV EPIC pn lightcurves with a binning of 50~s.
\label{fig:lightcurves-detail}}
\end{figure*}

Fig.~\ref{fig:lightcurves} shows 0.3--10~keV EPIC pn \src\ background subtracted lightcurves 
with a binning of 50~s for each of the observations listed in Table~\ref{tab:obslog}.
Despite the low count rate of the source in these observations, the narrow eclipses are clearly evident. Intervals used for spectral analysis (not affected by high background flares)
are marked on the top of each panel. We inspected the light curves together with the hardness ratio (counts in the 2--10~keV energy range divided by those between 0.3--2~keV) to search for variations in the out-of-eclipse emission, which could reveal residual accretion. For instance, detection of dipping activity would indicate the existence of absorption in an ionised atmosphere and therefore the existence of an accretion disc. Detailed plots of the light curves are shown in Fig.~\ref{fig:lightcurves-detail}. The eclipses ingresses and egresses are sharp. 

Outside of the eclipses the light curves do not show large deviations from the average count rate and the hardness ratio is consistent with being constant. As an example, the distribution of count rate between the first two eclipses in Obs~2 is consistent with a Poisson distribution of mean 0.54~s$^{-1}$ (\chisq\ of 21 for 13 degrees of freedom).

\subsection{X-ray spectra}
\label{sec:spectra}

For each observation, we first extracted EPIC and RGS spectra excluding the 
times of eclipses. We rebinned all the EPIC spectra to
over-sample the $FWHM$ of the energy resolution by a factor 3, 
and to have a minimum of 25 counts per
bin, to allow the use of the $\chi^2$ statistic. 
We used the RGS spectra with two different binnings: in 
Sections~\ref{sec:epic-rgs}-\ref{sec:sim-analysis} we rebinned
the RGS spectra to over-sample the $FWHM$ of the energy resolution by
a factor 3 and to have a minimum of 25 counts per bin, to be able to
consistently use the $\chi^2$ statistic for both EPIC and RGS
spectra. In Sect.~\ref{sec:rgs} we rebinned the RGS spectra with different
methods to be sensitive to narrow features (see below).
For the EPIC (RGS) spectra we used
the 0.3--10~keV (0.5-1.8~keV) energy band. Constant factors, fixed to 1 for the EPIC
pn spectrum, but allowed to vary for the EPIC MOS and RGS spectra, were
included multiplicatively in order to account for cross-calibration
uncertainties. 

We performed spectral analysis using XSPEC \citep{arnaud96conf}
version 12.6.0. We used the photo-electric cross-sections and abundances of
\citet{wilms00apj} to account for absorption by neutral gas (the {\tt tbabs} XSPEC model). 
Spectral uncertainties are given at 90\% confidence ($\Delta$\chisq = 2.71
for one interesting parameter) and upper limits at 95\% confidence.

\subsubsection{RGS spectral analysis}
\label{sec:rgs}

We examined the 0.5--1.8~keV (7--25~\ang) first and second order
RGS spectra to constrain the \nh\ in the direction of the source and to
search for the signature of narrow absorption and
emission features.  

We first rebinned the RGS spectra for each observation to over-sample the $FWHM$ of the energy resolution by
a factor 3 and to have a minimum of 10 counts per bin.

We could fit the RGS spectra of all the
observations with a continuum consisting of a 
neutron star atmosphere model component, suitable for cooling NSs (see Sect.~\ref{sec:epic-rgs}), modified by
photo-electric absorption from neutral material. 
We obtained a C-statistic \citep{cash79apj} of 283, 377, 612 and 561 for 242, 357, 591 and 530 degrees of freedom (d.o.f.) for Obs~1 to 4, respectively. The values of \nh\ in units of 10$^{21}$\,cm$^{-2}$ were 1.6\,$^{+0.4}_{-1.3}$, $<$\,0.7, $<$0.5 and $<$1.5 for Obs~1--4, respectively.
None of the observations showed 
narrow features. 

In order to increase the sensitivity to narrow lines and edges, we combined the RGS spectra of all the observations and rebinned the combined spectra to over-sample the $FWHM$ of the energy resolution by
a factor 3 and used the C-statistic. We could fit the spectra with the same model as the individual observations and obtained a C-statistic(d.o.f.) of 1651(1673) and \nh\ $<$\,0.5\,$\times$10$^{21}$\,cm$^{-2}$.

To evaluate the effect of the absorption component in our fit, we substituted the {\tt tbabs} component by the more recent {\tt tbnew} component and re-fitted the combined spectra. We obtained a C-statistic of 1651(1673) and \nh\,$<$\,0.4\,$\times$10$^{21}$\,cm$^{-2}$. 

Finally, we rebinned the merged spectra imposing a minimum signal-to-noise ratio (SNR) of 5 per channel, in order to minimise the background effects and re-fitted the first order spectra (the second order spectra do not have enough SNR to be fitted). We obtained a C-statistic of 115 (70) and \nh\,$<$\,0.3\,$\times$10$^{21}$\,cm$^{-2}$.

In summary, we obtained consistent values of \nh\ for different models of interstellar absorption and rebinning methods. 
However, we obtained significantly different values of \nh\ when using different continuum components. This is expected due to the combination of limited energy band of the RGS and low statistics of these observations. 
As an example, substituting the neutron star atmosphere component by a power-law component in the fits to the merged spectra we obtained values of \nh\ of 2.4\,$\pm$\,0.8 and $<$\,1.6\,$\times$10$^{21}$\,cm$^{-2}$ and C-statistic of 1647(1674) and 103(71) for a rebinning with oversampling by a factor 3 and with a minimum SNR of 5, respectively. Therefore, we conclude that the RGS spectra alone cannot constrain the value of \nh\ for these observations due to the poor statistics and the limited bandwidth, necessary to determine the continuum.

We did not find any significant narrow feature in the RGS merged spectra.

\subsubsection{EPIC and RGS spectral analysis}
\label{sec:epic-rgs}

We next fitted the EPIC and RGS spectra simultaneously for each observation with a continuum consisting of a
neutron star atmosphere component, to account for the thermal emission below $\sim$3~keV, and a power-law component, to account for emission above $\sim$3~keV, modified by photo-electric absorption from neutral material (model {\tt tbabs*(nsatmos+po)} in XSPEC). The {\tt nsatmos} model \citep{heinke06apj} includes a range of surface gravities and effective temperatures and incorporates thermal electron conduction and self-irradiation by photons from the compact object. It also assumes negligible ($<$10$^9$~G) magnetic fields and a pure hydrogen atmosphere.
The parameters of the {\tt nsatmos} component are the NS mass and radius, the effective temperature in the NS frame, the source distance and a normalisation factor, which parameterizes the fraction of the surface that is radiating. Throughout this work, we imposed a lower limit to the mass of the NS of 1.27\,$\Msun$, as derived by \citet{0748:bassa09mnras}. Further, since NSs in LMXBs do not show evidence, such as spin-powered pulsations or cyclotron spectral features, for a strong magnetic field, we kept the normalisation of the neutron star atmosphere model fixed to 1, which corresponds to the entire NS surface emitting. 

\begin{table*}[!ht]
\begin{center}
\caption[]{Best-fits to the 0.3--10~keV EPIC and 0.5--1.8~keV RGS spectra for each observation individually with the  {\tt tbabs*(nsatmos+po)} model. \kpl\ is the normalisation of the power-law component in units of 10$^{-6}$ \unitplnorm. F$_{tot}$ and F$_{pow}$ are the 0.3--10~keV total and power law unabsorbed fluxes in units of 10$^{-12}$ erg cm$^{-2}$ s$^{-1}$ and F$_{NS}^{bol}$ the 0.01--100~keV {\tt nsatmos} unabsorbed flux in the same units. The distance was fixed to 7.1~kpc and the {\tt nsatmos} normalisation to 1 during the fits (see text).}
\begin{tabular}{c@{\extracolsep{0.18cm}}c@{\extracolsep{0.18cm}}c@{\extracolsep{0.18cm}}c@{\extracolsep{0.18cm}}c@{\extracolsep{0.18cm}}c@{\extracolsep{0.18cm}}c@{\extracolsep{0.18cm}}c@{\extracolsep{0.18cm}}c@{\extracolsep{0.18cm}}c@{\extracolsep{0.18cm}}c}
\hline \noalign {\smallskip}
Observation & \nh\ & T$_{eff}$ & M & R & \phind\  & \kpl\ & F$_{tot}$ & F$_{pow}$ & F$_{NS}^{bol}$ & \rchisq\ (d.o.f.)  \\
\noalign {\smallskip}
   &  [10$^{21}$\,cm$^{-2}$] & [eV] & [M$_{\odot}$] & [km] & &  \\
\hline \noalign {\smallskip}
0560180701 & 0.70\,$\pm$\,0.11 & 118\,$\pm$\,4 & 1.4 (f) & 15.6\,$\pm$\,1.1 & 0.7\,$\pm$\,0.7 & 4\,$^{+10}_{-3}$ & 1.33\,$\pm$\,0.02 & 0.09\,$\pm$\,0.02 & 1.36\,$\pm$\,0.01 & 0.85 (247) \\
0560180701 & 0.73\,$\pm$\,0.10 &117\,$\pm$\,6 & 1.8 (f) & 15.1\,$^{+0.4}_{-1.2}$ & 0.7\,$^{+0.6}_{-0.8}$ & 4\,$^{+10}_{-3}$ & 1.34\,$\pm$\,0.02 & 0.09\,$\pm$\,0.02 & 1.37\,$\pm$\,0.02 & 0.85 (247) \\
0560180701 & 0.70\,$^{+0.13}_{-0.09}$ &  118\,$^{+119}_{-59}$ & $<$\,2.4 & 15.7\,$^{+1.0}_{-6.1}$ & 0.7\,$\pm$\,0.8 & 4\,$^{+10}_{-3}$ & 1.33\,$\pm$\,0.02 & 0.09\,$\pm$\,0.02 & 1.36\,$\pm$\,0.02 & 0.86 (246) \\
\noalign {\smallskip}

0605560401 & 0.76\,$\pm$\,0.09 & 112\,$\pm$\,3 & 1.4 (f) & 15.6\,$\pm$\,0.9 & -- & -- & 1.01\,$\pm$\,0.01 & -- & 1.12\,$\pm$\,0.01 & 1.04 (288) \\
0605560401 & 0.78\,$\pm$\,0.08 & 111\,$\pm$\,4 & 1.8 (f) & 15.1\,$\pm$\,0.9 & -- & -- & 1.02\,$\pm$\,0.01 & -- & 1.13\,$\pm$\,0.01 & 1.03 (288) \\
0605560401 & 0.77\,$\pm$\,0.09 &  111\,$^{+96}_{-51}$ & $<$\,2.4 & 15.2\,$^{+1.2}_{-6.5}$ & -- & -- & 1.02\,$\pm$\,0.01 & -- & 1.13\,$\pm$\,0.01 & 1.04 (287) \\
\noalign {\smallskip}

0605560501  & 0.48$\pm$0.07 & 114\,$\pm$\,3 & 1.4 (f)  & 13.1\,$\pm$\,0.7 & -- & -- & 0.833$\pm$0.008 & -- & 0.915$\pm$0.009 & 1.09 (404) \\
0605560501  & 0.50$\pm$0.07 & 114\,$\pm$\,6 & 1.8 (f)  & 12.1\,$\pm$\,1.0 & -- & -- & 0.839\,$\pm$\,0.008 &-- & 0.923\,$\pm$\,0.009& 1.09 (404) \\
0605560501  & 0.50$\pm$0.07 & 114\,$^{+89}_{-50}$ & $<$\,2.1 & 12.0\,$^{+1.8}_{-5.5}$ & -- & -- & 0.840$\pm$0.008 & -- & 0.923$\pm$0.009 & 1.09 (403) \\
\noalign {\smallskip}

0651690101 & 0.61\,$\pm$\,0.07 & 112\,$\pm$\,3 & 1.4 (f) & 14.0\,$\pm$\,0.7 & -- & -- & 0.831\,$\pm$\,0.008 & -- & 0.918\,$\pm$\,0.009 & 1.02 (466) \\
0651690101 & 0.64\,$\pm$\,0.07 & 110\,$\pm$\,5 & 1.8 (f) & 13.3\,$\pm$\,0.9 & -- & -- & 0.838\,$\pm$\,0.008 & -- & 0.928\,$\pm$\,0.009 & 1.02 (466) \\

0651690101 & 0.62\,$\pm$\,0.08 & 111\,$^{+100}_{-52}$ & $<$\,2.2 & 13.7\,$^{+1.0}_{-4.6}$& -- & -- & 0.833\,$\pm$\,0.008 & -- & 0.921\,$\pm$\,0.009 & 1.03 (465) \\

\noalign {\smallskip} \hline \label{tab:epic-rgs-fit}
\end{tabular}
\end{center}
\end{table*}

We performed 3 fits per observation: first we left the mass of the NS as a free parameter during the fit and then we fixed its value to the canonical mass of 1.4\,$\Msun$ and to 1.8\,$\Msun$ for comparison. We extrapolated the flux of the thermal component to the energy range of 0.01-100~keV to estimate the thermal bolometric flux.
The fits were acceptable, with reduced \chisq, \rchisq, $\sim$\,1 for 250--450 d.o.f., for all the observations. The inclusion of a power-law component improved the goodness of the fit only for Obs~1.
The best-fit results are shown in Table~\ref{tab:epic-rgs-fit} for a distance of 7.1~kpc. The errors on the temperature and radius of the NS are significantly smaller when the mass of the NS is fixed. The value of \nh\ shows significant differences among observations. This clearly affects the accuracy with which we can measure the value of the NS radius and its effective temperature, and shows that a single observation is not sufficient to impose constraints on the mass and radius of the NS. 
   
\subsubsection{Simultaneous fits for all the datasets}
\label{sec:sim-analysis}

\begin{figure*}
\centerline{
\hspace{-0.6cm}
\includegraphics[angle=0,width=0.53\textwidth]{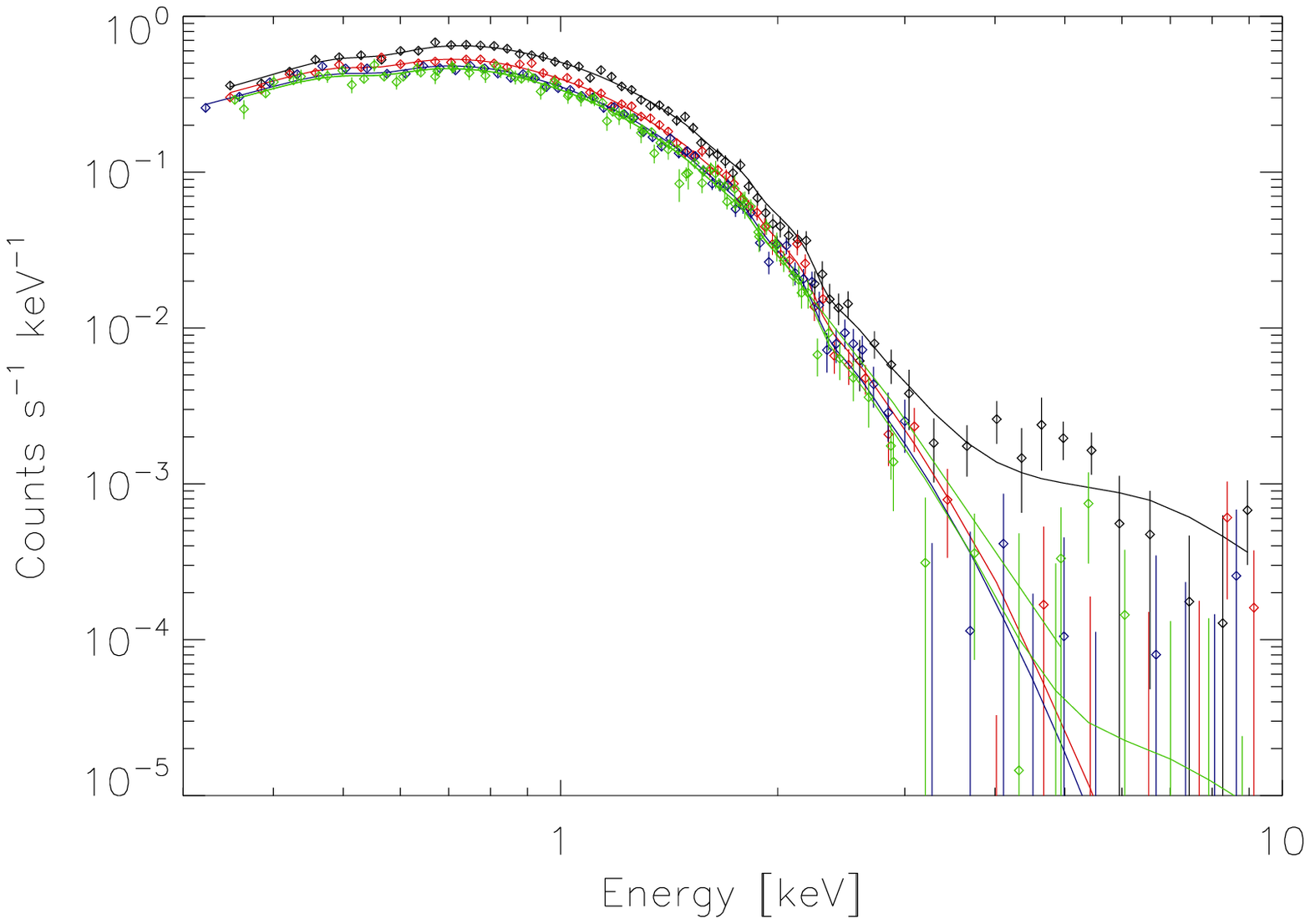}
\hspace{-0.3cm}
\includegraphics[angle=0,width=0.53\textwidth]{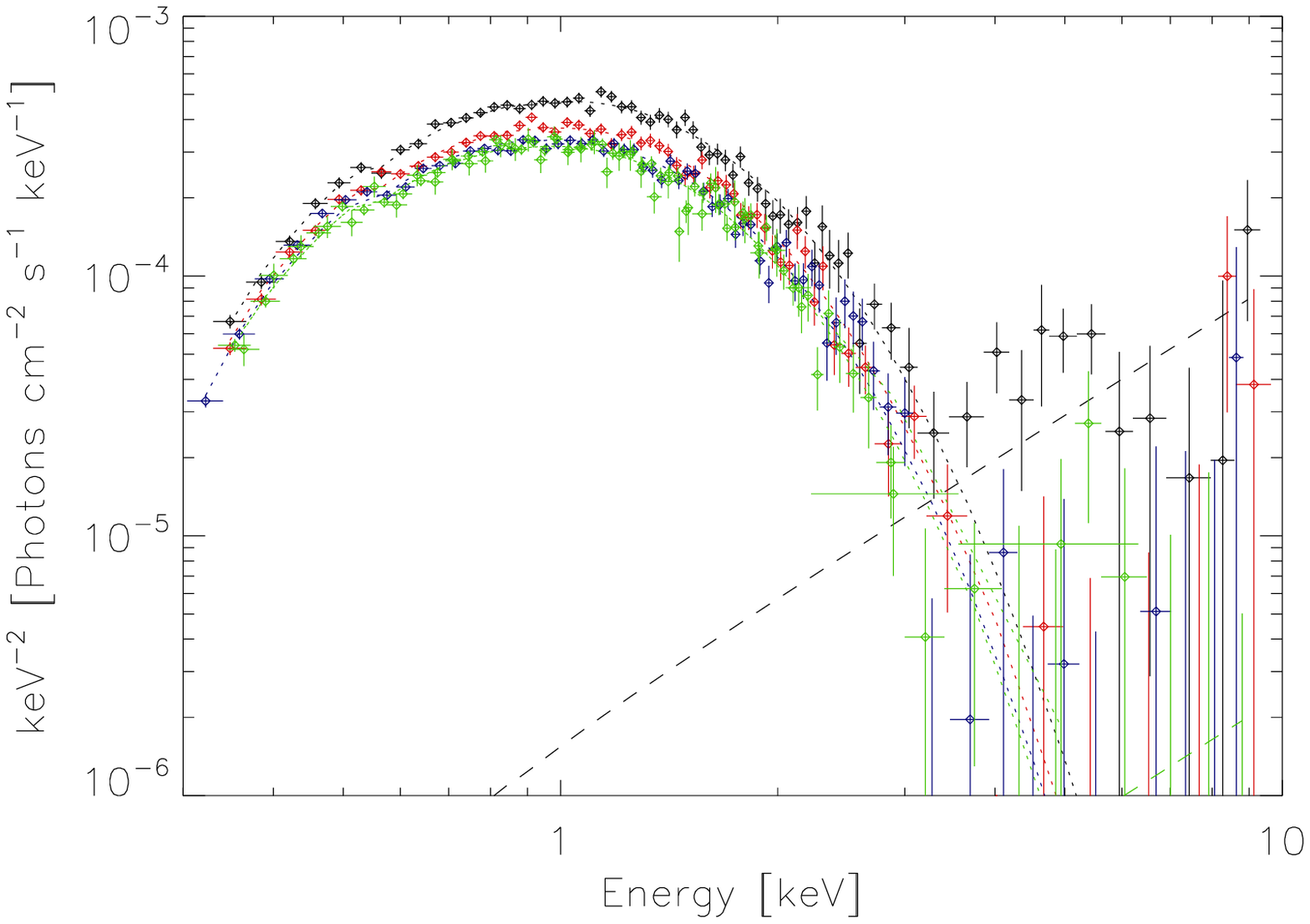}
}
\caption{{\it Left}: EPIC pn spectra fitted with the model {\tt tbabs*(nsatmos+po)}. Obs~1--4 are shown in black, red, blue and green, respectively. 
{\it Right}: Unfolded EPIC pn spectra. The colors are the same as in the left panel.
\label{fig:fits}}
\end{figure*}

In order to obtain the best possible constraints for the mass and radius of the NS with the \xmm\ set of observations, we next fitted the EPIC and RGS spectra of all the observations simultaneously. We first assumed that the value of the absorption does not change among observations and tied the values of \nh, and mass and radius of the NS for all the observations. Since the contribution of the power-law component was only significant in the first observation (see Sect.~\ref{sec:epic-rgs}), we also tied the index of the power-law component for the simultaneous fit and allowed the normalisation of this component to vary from one observation to another. Similarly, we allowed the temperature of the NS atmosphere component to vary between observations to account for the expected cooling of the NS crust. Table~\ref{tab:fit-sim} shows the results of the best-fit model for 3 values of the distance: 5.9, 7.1 and 8.3~kpc. 
The fits are acceptable with a \rchisq\ of 1.03 for 1419 d.o.f.  
Fig.~\ref{fig:fits} shows the results of the fits for a distance of 
7.1~kpc (note that we only show the EPIC pn spectra for clarity, but the fit included the EPIC MOS1 and MOS2 and the RGSs spectra too).
Fits of the thermal component to a blackbody model yielded similar \rchisq, but 
implied an emitting area much smaller than a neutron star's surface, in agreement with the expectations (since the
emergent spectra at T$_{eff}$\,$\approxlt$\,5\,$\times$\,10$^6$\,K is very different from a blackbody, \citet{brown98apjl}). Therefore, we do not discuss the blackbody fits further. Substituting the power-law component by a thermal Comptonisation component (model {\tt comptt} in XSPEC) yielded a similar \rchisq\ and the parameters of the NS atmosphere were not affected, i.e. the values obtained for the mass, radius and temperature of the NS are robust against changes of the model for the emission component above $\sim$3~keV.

\begin{figure*}
\centerline{\includegraphics[angle=0,width=0.39\textwidth]{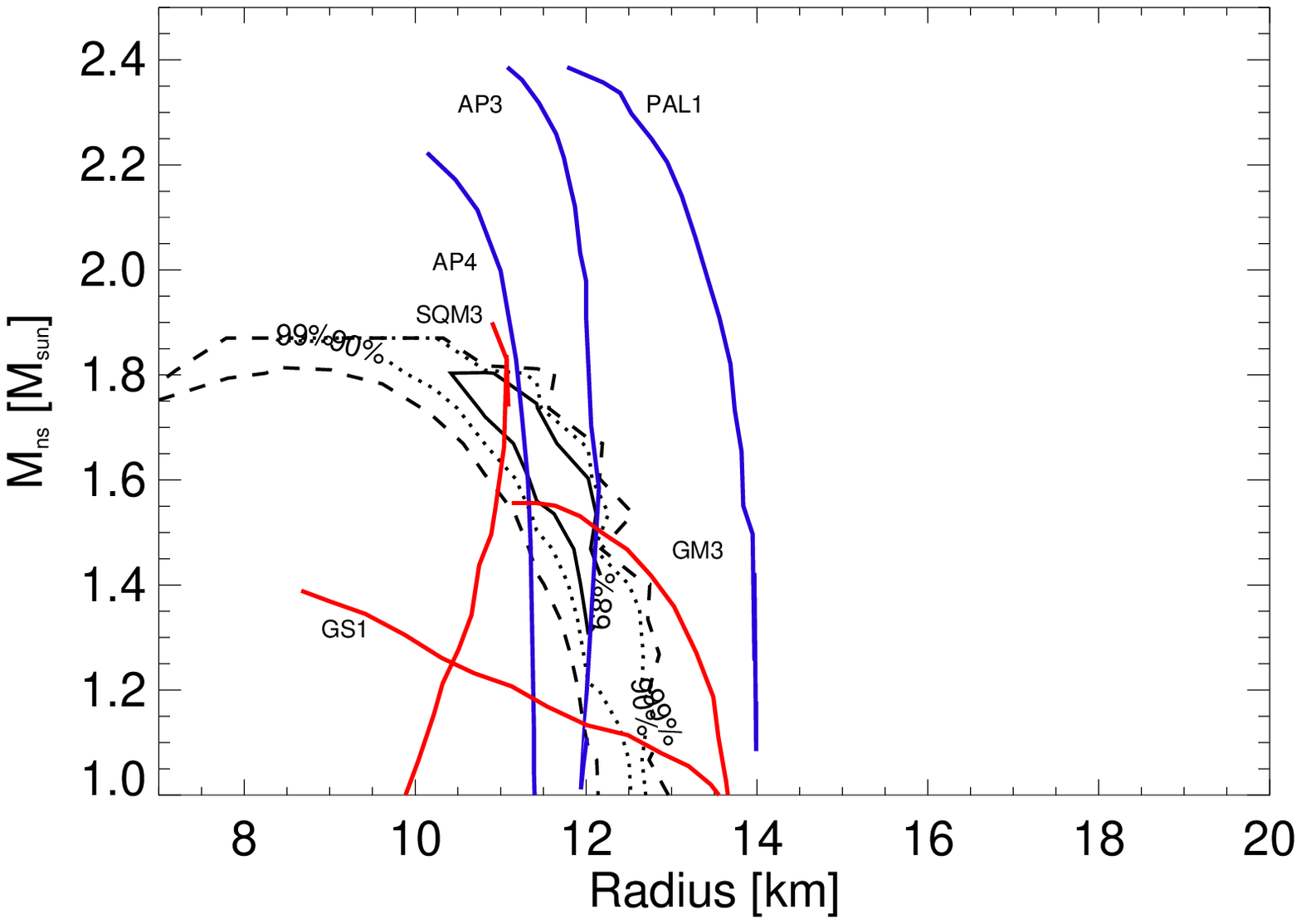}
\hspace{-0.5cm}
\includegraphics[angle=0,width=0.39\textwidth]{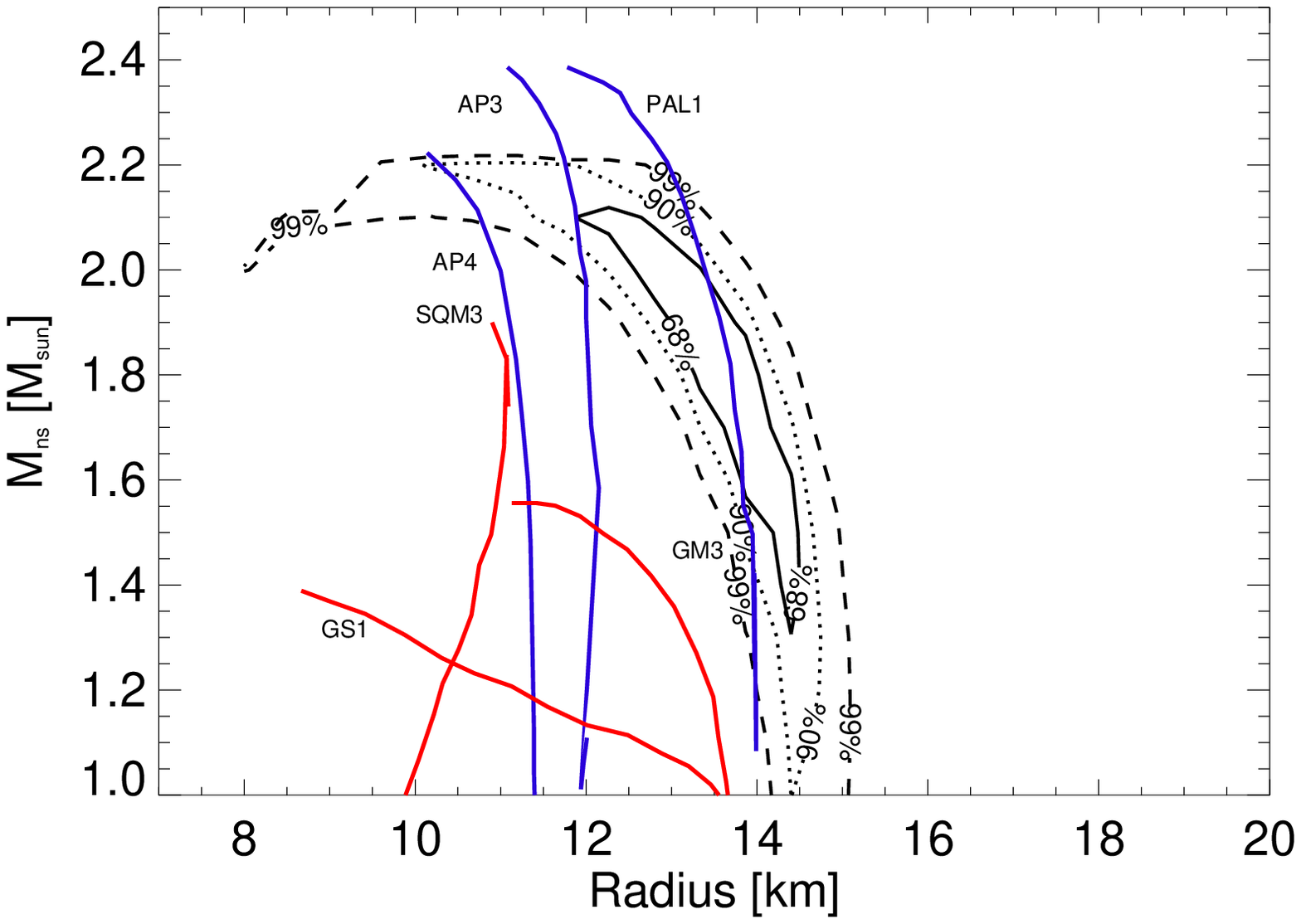}
\hspace{-0.5cm}
\includegraphics[angle=0,width=0.39\textwidth]{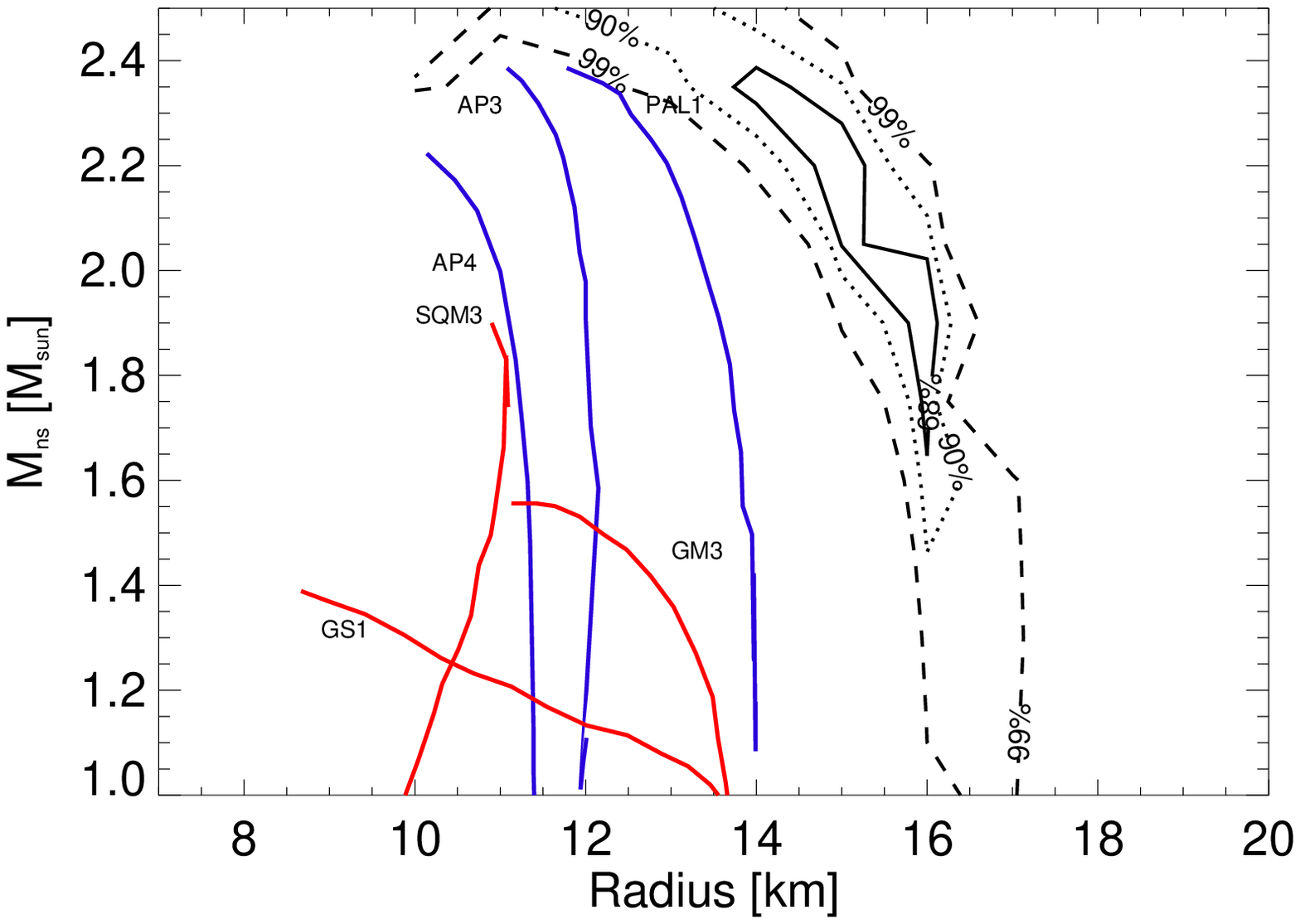}
}
\caption{Contour plots for the mass and radius of the NS for the fits shown in Table~\ref{tab:fit-sim} for a distance of 5.9 (left), 7.1 (centre) and 8.3 (right) kpc. The solid, dotted and dashed lines represent the 68, 90 and 99\% probability contours. Examples of mass and radius predictions for representative equations of state with neutron and protons (blue lines) and including quarks, hyperons or kaons (red lines) 
are shown for comparison \citep[see][]{lattimer01apj,lattimer07pr}.
\label{fig:contours}}
\end{figure*}

Fig.~\ref{fig:contours} shows the contour plots obtained for the mass and radius of the NS for the three distances considered in Table~\ref{tab:fit-sim}. As we increase the distance of the source in the model, both the best-fit values of the mass and the radius and the allowed region for these parameters increase.

\begin{table*} [!ht]
\begin{center}
\caption[]{Best-fits to the 0.3--10~keV EPIC and 0.5--1.8~keV RGS spectra for all the observations simultaneously with the {\tt tbabs*(nsatmos+po)} model (see caption of Table~\ref{tab:epic-rgs-fit} for definitions). 
}
\begin{tabular}{c@{\extracolsep{0.18cm}}c@{\extracolsep{0.18cm}}c@{\extracolsep{0.18cm}}c@{\extracolsep{0.18cm}}c@{\extracolsep{0.18cm}}c@{\extracolsep{0.18cm}}c@{\extracolsep{0.18cm}}c@{\extracolsep{0.18cm}}c@{\extracolsep{0.18cm}}c@{\extracolsep{0.18cm}}c@{\extracolsep{0.18cm}}}
\hline \noalign {\smallskip}
Observation & \nh\ & kT$_{eff}^{\infty}$ & M & R & \phind\  & \kpl\ & F$_{tot}$ & F$_{pow}$ & F$_{NS}^{bol}$ & \rchisq\ (d.o.f.)  \\
\noalign {\smallskip}
   &  [10$^{21}$\,cm$^{-2}$] & [eV] & [M$_{\odot}$] & [km] & &  \\
\hline \noalign {\smallskip}
\multicolumn{10}{c}{Distance: 5.9~kpc} \\
\noalign {\smallskip}
0560180701 & 0.66\,$\pm$\,0.04 & 118.2\,$^{+26}_{-21}$ & 1.52\,$\pm$\,0.5 & 11.8\,$^{+0.7}_{-2.2}$ & 0.24\,$\pm$\,0.7 & 1.7\,$^{+3.3}_{-1.2}$ & 1.33\,$\pm$\,0.01 & 0.09\,$\pm$\,0.02 & 1.36\,$\pm$\,0.01 & 1.03 (1419) \\
0605560401 & & 111.9\,$^{+25}_{-20}$ & & & & $<$\,0.27 & 0.98\,$\pm$\,0.01 & $<$\,0.011 & 1.08\,$\pm$\,0.01 & \\
0605560501 &  & 109.1\,$^{+24}_{-20}$ & & & & $<$\,0.24 & 0.885\,$\pm$\,0.009 & $<$\,0.007 & 0.98\,$\pm$\,0.01 & \\
0651690101 & & 107.6\,$^{+24}_{-20}$ & & & & $<$\,0.43 & 0.841\,$\pm$\,0.008 & $<$\,0.014 & 0.933\,$\pm$\,0.009 & \\
\noalign {\smallskip}
\multicolumn{10}{c}{Distance: 7.1~kpc} \\
\noalign {\smallskip}
0560180701 & 0.64\,$\pm$\,0.04 & 120.1\,$^{+28}_{-23}$ & 1.78\,$^{+0.4}_{-0.6}$ & 13.7\,$^{+1.0}_{-2.7}$ & 0.23\,$\pm$\,0.7 & 1.7\,$^{+3.3}_{-1.2}$ & 1.32\,$\pm$\,0.01 & 0.09\,$\pm$\,0.02 & 1.35\,$\pm$\,0.01 & 1.03 (1419) \\
0605560401 & & 113.6\,$^{+26}_{-21}$ & & & & $<$\,0.26 & 0.98\,$\pm$\,0.01 &  $<$\,0.011 & 1.08\,$\pm$\,0.01 & \\
0605560501 &  &110.8\,$\pm$\,14 & & & & $<$\,0.23 & 0.880\,$\pm$\,0.009 & $<$\,0.007  & 0.97\,$\pm$\,0.01 &\\
0651690101 & & 109.5\,$^{+26}_{-21}$   & & & & $<$\,0.42 & 0.834\,$\pm$\,0.008 &  $<$\,0.014 & 0.925\,$\pm$\,0.009 & \\
\noalign {\smallskip}
\multicolumn{10}{c}{Distance: 8.3~kpc} \\
\noalign {\smallskip}
0560180701 & 0.63\,$\pm$\,0.04 & 121.4\,$^{+28}_{-23}$ & 2.12\,$^{+0.4}_{-0.8}$ & 15.2\,$^{+1.5}_{-3.0}$ & 0.28\,$^{+0.6}_{-0.8}$ & 1.8\,$^{+3.0}_{-1.3}$ & 1.31\,$\pm$\,0.01 & 0.08\,$\pm$\,0.02 & 1.34\,$\pm$\,0.01 & 1.03 (1419) \\
0605560401 & & 114.9\,$^{+20}_{-17}$ & & & & $<$\,0.38& 0.97\,$\pm$\,0.01 & $<$\,0.011 & 1.07\,$\pm$\,0.01 & \\
0605560501 &  & 112.0\,$^{+25}_{-20}$ & & & & $<$\,0.21 & 0.876\,$\pm$\,0.009 & $<$\,0.007 & 0.97\,$\pm$\,0.01 & \\
0651690101 & & 110.7\,$\pm$\,18 & & & & $<$\,0.40 & 0.832\,$\pm$\,0.008 & $<$\,0.014 & 0.919\,$\pm$\,0.009 & \\
\noalign {\smallskip} \hline \label{tab:fit-sim}
\end{tabular}
\end{center}
\end{table*}

Next, we investigated how the errors of the fit vary when some parameters are fixed or untied in the fit. First, since we have indications from the individual fits performed in Sect.~\ref{sec:epic-rgs} that the value of \nh\ may be different among observations, we allowed this parameter to vary in the fits (Case 1 in Table~\ref{tab:epic-all-trials}).
Second, since the index of the power-law component is not well constrained, we fitted the spectra fixing the index to the best-fit value ($\Gamma$\,=\,0.24) and to 1 (Cases 2 and 3 in Table~\ref{tab:epic-all-trials}). Finally, we fixed the mass to the best-fit value of 1.78~$\Msun$ and to a more canonical value of 1.40~$\Msun$, but allowed to vary the index of the power law (Cases 4 and 5 in Table~\ref{tab:epic-all-trials}). 

\begin{table*}[!ht] 
\begin{center}
\caption[]{Best-fits to the 0.3--10~keV EPIC and 0.5--1.8~keV RGS spectra for all the observations simultaneously with the  {\tt tbabs*(nsatmos+po)} model (see caption of Table~\ref{tab:epic-rgs-fit} for definitions).  
}
\begin{tabular}{c@{\extracolsep{0.18cm}}c@{\extracolsep{0.18cm}}c@{\extracolsep{0.18cm}}c@{\extracolsep{0.18cm}}c@{\extracolsep{0.18cm}}c@{\extracolsep{0.18cm}}c@{\extracolsep{0.18cm}}c@{\extracolsep{0.18cm}}c@{\extracolsep{0.18cm}}c@{\extracolsep{0.18cm}}c@{\extracolsep{0.18cm}}}
\hline \noalign {\smallskip}
Observation & \nh\ & kT$_{eff}^{\infty}$ & M & R & \phind\  & \kpl\ & F$_{tot}$ & F$_{pow}$ & F$_{NS}^{bol}$ & \rchisq\ (d.o.f.)  \\
\noalign {\smallskip}
   &  [10$^{21}$\,cm$^{-2}$] & [eV] & [M$_{\odot}$] & [km] & &  \\
\hline \noalign {\smallskip}
\noalign {\smallskip}
\multicolumn{10}{c}{Case 1:} \\
0560180701 & 0.66\,$^{+0.06}_{-0.03}$ & 120.0$^{+36}_{-27}$ & 1.77$^{+0.4}_{-0.7}$ & 13.7$^{+1.5}_{-3.2}$ & 0.26\,$\pm$\,0.7 & 1.8$^{+3.6}_{-1.2}$ & 1.32\,$\pm$\,0.01 & 0.09\,$\pm$\,0.02 & 1.34\,$\pm$\,0.01 & 1.03 (1416) \\
0605560401 & 0.67\,$\pm$\,0.06 & 113.8\,$^{+30}_{-24}$ & & & & $<$0.25 & 0.99\,$\pm$\,0.01 &  $<$\,0.011 & 1.09\,$\pm$\,0.01  & \\
0605560501 &  0.60\,$\pm$\,0.05 & 110.5\,$^{+32}_{-25}$& & & & $<$0.31 & 0.869\,$\pm$\,0.009 & $<$\,0.007 & 0.96\,$\pm$\,0.01  &\\
0651690101 & 0.66\,$\pm$\,0.06 & 109.5\,$^{+32}_{-25}$ & & & & $<$0.41 & 0.840\,$\pm$\,0.008 & $<$\,0.014 &0.930\,$\pm$\,0.009 &\\
\hline\noalign {\smallskip}
\multicolumn{10}{c}{Case 2: } \\
0560180701 & 0.64\,$\pm$\,0.04 & 120.1\,$\pm$\,16  & 1.78\,$^{+0.6}_{-0.1}$ & 13.7\,$^{+0.3}_{-1.5}$ & 0.24(f) & 1.7\,$^{+2.1}_{-1.3}$ & 1.32\,$\pm$\,0.01 & 0.09\,$\pm$\,0.02 & 1.35\,$\pm$\,0.01 & 1.03 (1420) \\
0605560401 & & 113.6\,$^{+20}_{-17}$ & & & & $<$\,0.21 & 0.98\,$\pm$\,0.01 & $<$\,0.011 & 1.08\,$\pm$\,0.01 & \\
0605560501 &  & 110.8\,$\pm$\,15 & & & & $<$\,0.14 & 0.880\,$\pm$\,0.009 & $<$\,0.007 & 0.97\,$\pm$\,0.01 &\\
0651690101 & & 109.3\,$^{+25}_{-21}$ & & & & $<$\,0.27 & 0.836\,$\pm$\,0.008 & $<$\,0.014 & 0.925\,$\pm$\,0.009 & \\
\hline\noalign {\smallskip}
\multicolumn{10}{c}{Case 3: } \\
0560180701 & 0.65\,$\pm$\,0.04 & 119.6\,$\pm$\,16 & 1.79\,$^{+0.2}_{-0.5}$ & 13.8\,$^{+0.2}_{-1.5}$ & 1 (f) & 5.3\,$\pm$\,1.3 & 1.31\,$\pm$\,0.01 & 0.08\,$\pm$\,0.02 & 1.34\,$\pm$\,0.01 & 1.04 (1420) \\
0605560401 & & 113.4\,$\pm$\,16  & & & & $<$\,0.55 & 0.98\,$\pm$\,0.01 & $<$\,0.009 & 1.08\,$\pm$\,0.01 & \\
0605560501 &  & 110.6\,$^{+23}_{-19}$  & & & & $<$\,0.67 & 0.881\,$\pm$\,0.009 & $<$\,0.010 & 0.97\,$\pm$\,0.01 &\\
0651690101 & & 109.1\,$^{+25}_{-20}$ & & & & $<$\,0.85 & 0.837\,$\pm$\,0.008 & $<$\,0.013 & 0.927\,$\pm$\,0.009 & \\
\hline\noalign {\smallskip}
\multicolumn{10}{c}{Case 4: } \\
0560180701 & 0.64\,$\pm$\,0.04 & 119.8\,$\pm$\,2.5 & 1.78(f) & 13.7\,$\pm$\,0.5 & 0.24\,$\pm$\,0.7 & 1.7\,$^{+3.3}_{-1.2}$ & 1.32\,$\pm$\,0.01 & 0.09\,$\pm$\,0.02 & 1.35\,$\pm$\,0.02 & 1.03 (1420) \\
0605560401 & & 113.4\,$\pm$\,2.1 & & & & $<$\,0.26 & 0.98\,$\pm$\,0.01 & $<$\,0.011 & 1.08\,$\pm$\,0.01 & \\
0605560501 & & 110.5\,$\pm$\,2.0 & & & & $<$\,0.23 & 0.879\,$\pm$\,0.009 & $<$\,0.007 & 0.97\,$\pm$\,0.01 &\\ 
0651690101 & & 109.3\,$\pm$\,2.0 & & & & $<$\,0.42 & 0.836\,$\pm$\,0.008 & $<$\,0.014 & 0.925\,$\pm$\,0.009 & \\
\hline\noalign {\smallskip}
\multicolumn{10}{c}{Case 5: } \\
0560180701 & 0.62\,$\pm$\,0.04 & 121.3\,$\pm$\,1.7 & 1.40(f) & 14.4\,$\pm$\,0.4 & 0.21\,$\pm$\,0.7 & 1.6\,$^{+3.1}_{-1.1}$ & 1.31\,$\pm$\,0.01 & 0.09\,$\pm$\,0.02 & 1.34\,$\pm$\,0.01 & 1.03 (1420) \\
0605560401 & & 114.8\,$\pm$\,1.6 & & & & $<$\,0.25 & 0.97\,$\pm$\,0.01 & $<$\,0.011 & 1.07\,$\pm$\,0.01 & \\
0605560501 &  & 111.9\,$\pm$\,1.5 & & & & $<$\,0.21 & 0.874\,$\pm$\,0.009 & $<$\,0.007 & 0.96\,$\pm$\,0.01 &\\
0651690101 & & 110.7\,$\pm$\,1.5 & & & & $<$\,0.40 & 0.831\,$\pm$\,0.008 & $<$\,0.014 & 0.92\,$\pm$\,0.009 & \\
\hline\noalign {\smallskip}
\end{tabular}
\label{tab:epic-all-trials}
\end{center}
\end{table*}

The \nh\ adopts slightly different values when allowed to vary among observations. However, this small change of \nh\ does not influence the values of the other parameters significantly. In contrast, Table~\ref{tab:epic-all-trials} shows that the large uncertainty in the mass of the NS drives the uncertainties of the other parameters of the fit to unrealistically large values. For example, we detect variations in the effective temperature of the NS atmosphere between observations as small as 1~eV, but the calculated uncertainty is more than an order of magnitude larger. The large uncertainty in the mass arises from the fact that there are three parameters required to specify a NS model fully: its mass, radius and effective temperature \citep{heinke06apj}. Therefore, for a given effective temperature, there are several acceptable pairs M$_{NS}$-R$_{NS}$ \citep[see Appendix B of][for details]{heinke06apj}. The uncertainty in the index of the power-law component does not have a significant effect in the uncertainties of the other parameters. Therefore, in the studies of the variations of the temperature of the NS atmosphere among observations that follow we took into account the errors listed in Table~\ref{tab:epic-all-trials} for Case 4. We also observe that changing the index of the power-law component from the best-fit value 0.24 to 1 has the effect of increasing significantly the normalisation of this component, although the contribution to the total flux remains unchanged. 

\begin{figure*}[!ht]
\centerline{
\hspace{-0.5cm}
\includegraphics[angle=0,width=0.55\textwidth]{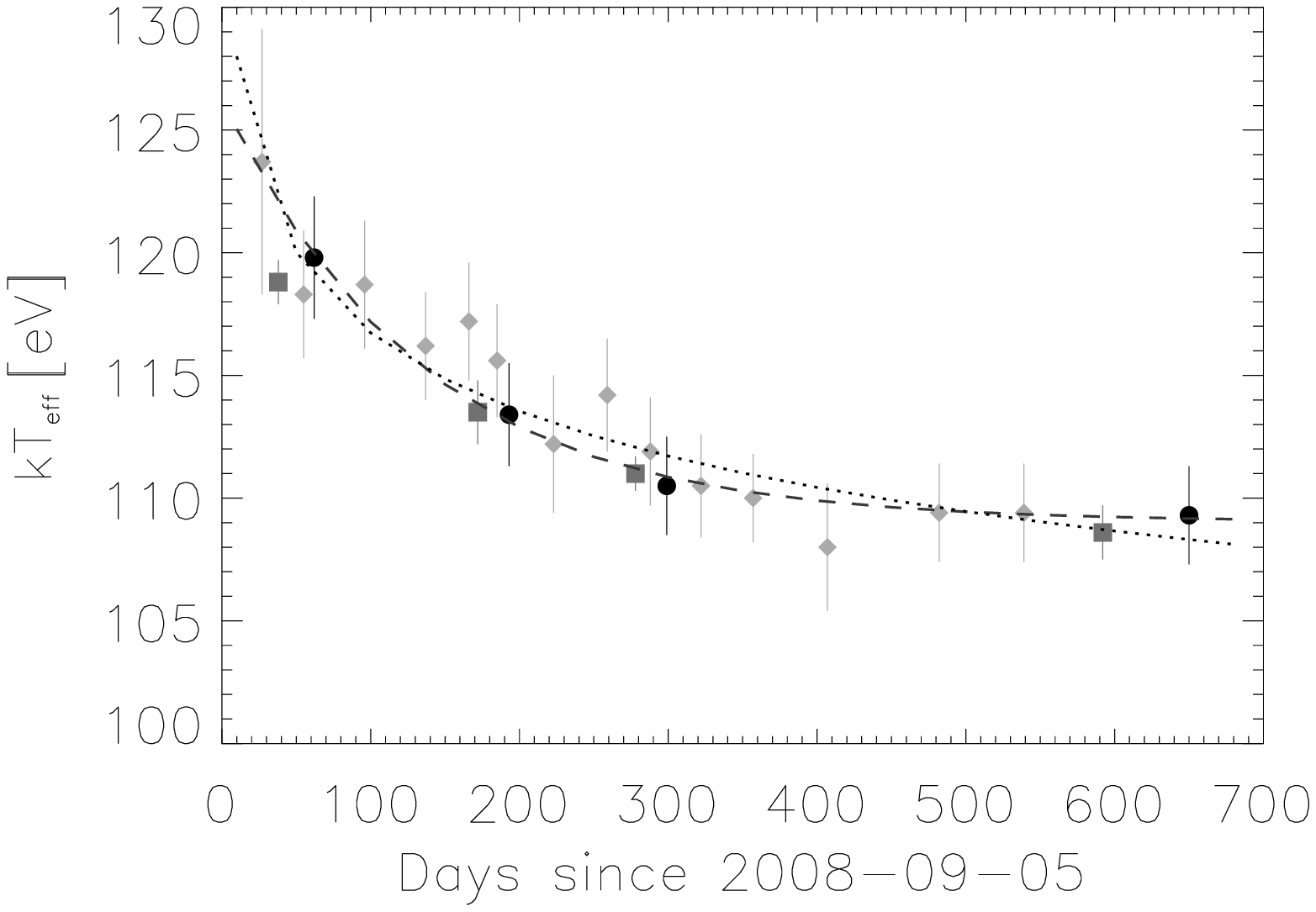}
\hspace{-0.5cm}
\includegraphics[angle=0,width=0.55\textwidth]{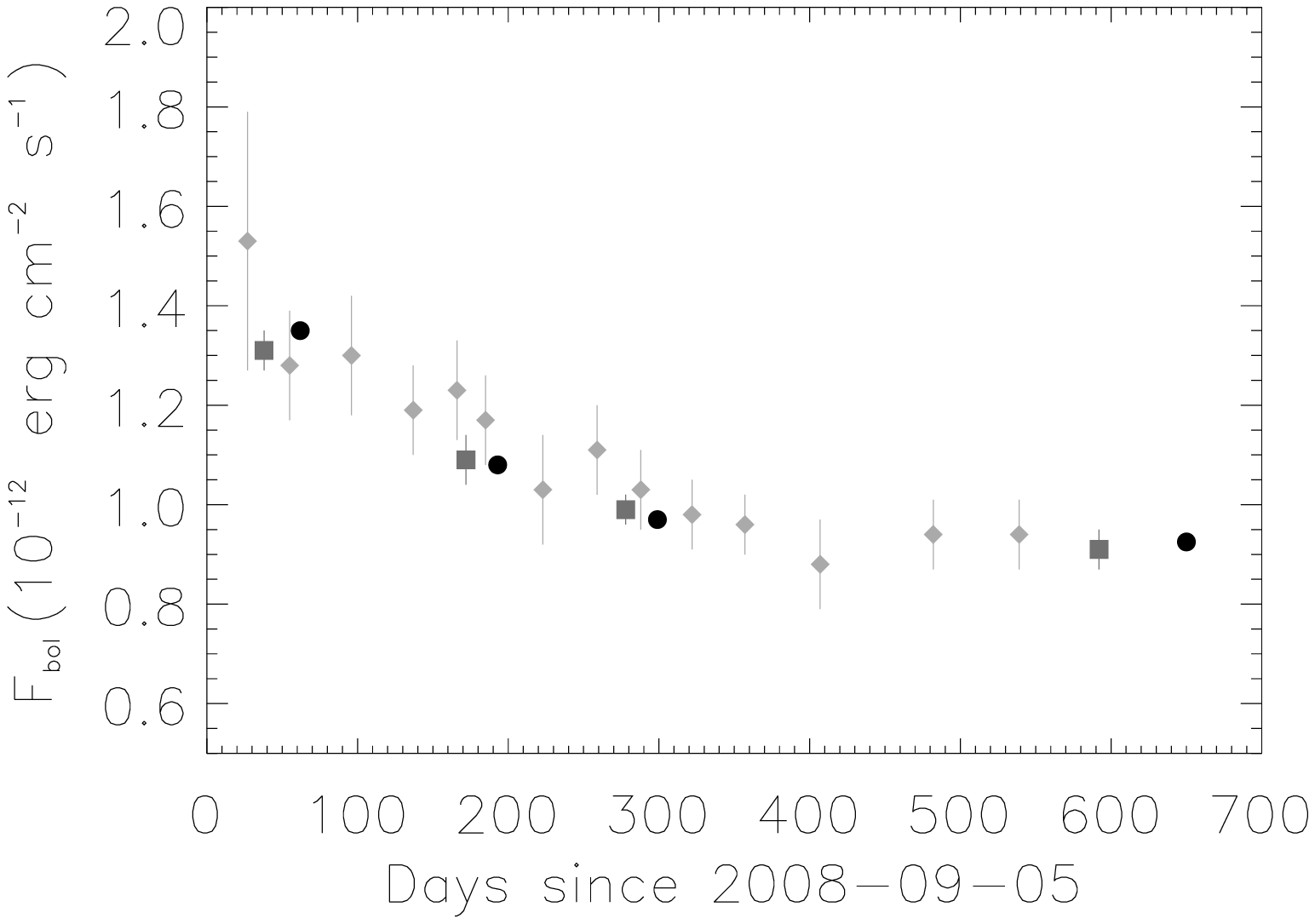}
}
\caption{Decay of effective temperature (left panel) and thermal bolometric flux (right panel) with time for the \xmm\ observations of \src. The dashed and dotted lines show the fit to a exponential function and to a power law, respectively. The \xmm, $Chandra$ and $Swift$ points are represented with black circles, dark grey squares and light grey diamonds, respectively. The \xmm\ points are the values obtained for Case~4 in Table~\ref{tab:epic-all-trials}. The $Chandra$ and $Swift$ points have been taken from \citet{0748:degenaar10mnras}.
\label{fig:decay-curve}}
\end{figure*}

We investigated the decay shape of the thermal flux and the temperature of the NS atmosphere as inferred from the {\tt nsatmos} parameters. We fitted the temperature curve with two models: an exponential decay function of the form $y(t)=a e^{-(t-t_0)/\tau}+b$, where $a$ is a normalisation constant, $t_0$ the start time of the cooling curve and $\tau$ the e-folding time (timescale for the temperature to decay by a factor of e$^{-1}$), and a power-law component of the form $y(t)=A(t-t_0)^B$. Following \citet{0748:degenaar10mnras}, we fixed $t_0$ to 2008 September 5, which is between the first non-detection by RXTE/PCA and the first $Swift$ observation of the source. Taking into account only the \xmm\ observations presented here, we found an e-folding time of 133.5\,$\pm$\,87.8~d, a normalisation of  17.2\,$\pm$\,5.8~eV and a constant level of 109.1\,$\pm$\,2.2 for the exponential fit (\rchisq\ of 0.06 for 1 d.o.f.). When fitting the decay curve with a power-law, we found parameter values of A\,=\,141.0\,$\pm$\,8.4~eV and B\,=\,-0.04\,$\pm$\,0.01 (\rchisq\ of 0.4 for 2 d.o.f.). The results of these fits are shown in Fig.~\ref{fig:decay-curve}. We note that  with only four points we cannot favour one of the above fits. However, the very small change of temperature and flux between Obs~3 and 4 (less than 2 and 6\% respectively) could point to a flattening of the cooling curve. If this is confirmed with future observations, the current power-law fit could be discarded in favour of a broken power-law fit with a tentative break between days 200 and 300. 

We next fitted the temperature curve taking into account the \xmm, $Chandra$ and $Swift$ values simultaneously. We found an e-folding time of 220\,$\pm$\,65~d, a normalisation of  14.0\,$\pm$\,1.4~eV and a constant level of 107.6\,$\pm$\,1.5~eV for the exponential fit (\rchisq\ of 0.39 for 19 d.o.f.). When fitting the decay curve with a power-law, we found parameter values of A\,=\,135.8\,$\pm$\,2.5~eV and B\,=\,-0.035\,$\pm$\,0.003 (\rchisq\ of 0.51 for 20 d.o.f.). In summary, the best-fit values obtained when $Chandra$ and $Swift$ values are taken into account are consistent within the errors with the values obtained only with the \xmm\ values. We note that we took the $Chandra$ and $Swift$ values from \citet{0748:degenaar10mnras}, who fixed the mass of the NS to 1.4~$\Msun$ and the distance to \src\ to 7.4~kpc in their fits. Although we fixed the mass of the NS to its best-fit value of 1.78~$\Msun$ and the distance to 7.1~kpc to obtain the \xmm\ points, Table~\ref{tab:epic-all-trials} shows that the temperature values are consistent within the errors for a mass of 1.78~$\Msun$ (Case 4) and 1.4~$\Msun$ (Case 5). In addition, Table~\ref{tab:fit-sim} shows that increasing the distance from 7.1 to 8.3~kpc causes significant changes in the values of the mass and radius of the NS but not in the effective temperature. This explains that we obtain consistent decay curves when the $Chandra$ and $Swift$ values are included in the fit. We attribute the small errors of the $Chandra$ points compared to the \xmm\ points (despite the poorer quality of the $Chandra$ spectra) to the fact that \citet{0748:degenaar10mnras} fixed all the parameters of the fit except the effective temperature and the normalisation of the power-law component. Therefore, the added value of the $Chandra$ and $Swift$ points should be taken with caution.
 
\subsection{OM data}
\label{sec:om}

We extracted images from the UVW1, UVM2 and UVW2 filter OM exposures for Obs~1. Based on the expected decreasing optical magnitude of the source, we chose to perform Obs~2--4 only with the U filter to maximise the sensitivity. OM data were not available for Obs~3 due to a technical error. For Obs~2 and 4 we extracted images and light curves for all OM exposures. 

The images show a source consistent with the position of \src\ which fades from Obs~1 to 4. In Obs~1, we obtained an average magnitude of 18.4\,$\pm$\,0.1 in the UVW1 filter. The source was not detected in the UVM2 or UVW2 filters. For Obs~2 and 4 we obtained average U optical magnitudes of 21.7\,$\pm$\,1.8 and 22.3\,$\pm$\,3.2, respectively. We note that \src\ is not detected in 4 out of 10 exposures in Obs~2 and in 14 out of 19 exposures in Obs~4.

\begin{figure}
\centerline{
\hspace{-0.5cm}
\includegraphics[angle=0,width=0.53\textwidth]{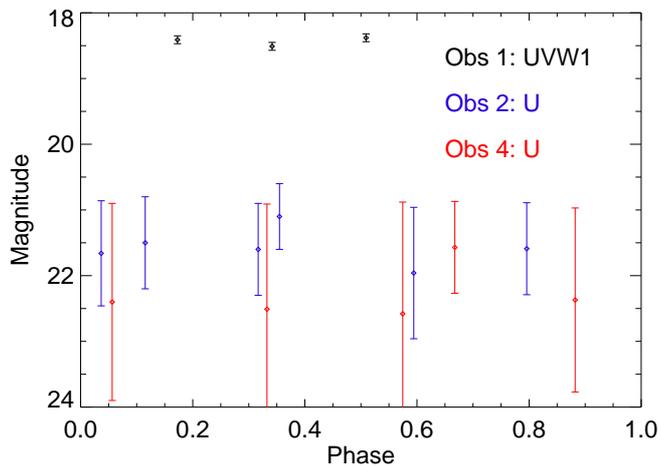}
}
\caption{Detections of \src\ with the OM monitor as a function of phase.
\label{fig:om}}
\end{figure}

Fig.~\ref{fig:om} shows the detections in all observations as a function of phase. Clearly, the optical magnitude was still decreasing in Obs~4 compared to Obs~2.  Unfortunately the errors are too large since the source is falling below detectability with OM. Therefore, we cannot determine the presence of a modulation with orbital phase.

\section{Discussion}

We analysed four \xmm\ observations of \src\ spanning 19 months and starting in November 2008,
when a significant halt of accretion into the NS was detected.

The light curves show a persistent level of emission without significant variations of intensity, except for the eclipse periods, which are clearly seen in all the observations. The persistent count rate decreased steadily by 
40\% from the first to the fourth observation.

All the observations show a soft spectral component between 0.3 and $\sim$3~keV,  which we fitted with a NS atmosphere model. The bolometric flux of the thermal component decreased by 30\% between the first and the last observation. In parallel, we observe a 10\% decrease in the effective surface temperature of the NS from 120 to 109~eV, which we interpret as the cooling of the NS crust towards thermal equilibrium with the core, after having been heated by accretion during the 24~year outburst. We detected only a 5\% decrease in flux and a 1.2\% decrease in temperature  between the last two observations, indicating that the NS crust may be close to reach thermal equilibrium with the core. Fitting the inferred temperatures with an exponential decay plus a constant offset yields an e-folding time of 133\,$\pm$\,88~days.

In addition to the thermal component, the first observation shows a component above $\sim$2-3~keV that we fitted with a power-law component, and which represents $\sim$\,7\% of the total 0.3-10~keV flux. The index of the power law is poorly constrained, with a value of 0.24$\pm$0.7. This value is much lower than the one found by \citet{0748:degenaar10mnras} in a previous analysis of this observation, 1.7$\pm$0.5, but the flux contribution to the total flux is similar. The index found in Obs~1 is consistent within the errors with values found for power-law components in other cooling NSs \citep[e.g.][]{cenx4:cackett10apj}. Although the origin of this component is still unknown, it has been suggested that residual low-level accretion onto the magnetosphere or a shock from a pulsar wind could account for the flux levels observed \citep{campana98aar}. The fact that we do not detect dips in the light curves indicates that residual accretion onto the NS via an accretion disc is unlikely. 

We do not observe any significant contribution of the power-law component after 6 November 2008. We determined upper limits to the contribution of this component to the total flux of 1.1, 0.8 and 1.7\% in the \xmm\ observations on 17 March and 1 July 2009 and  17 June 2010, respectively. In contrast, \citet{0748:degenaar10mnras} report a significant changing contribution between 5 and 15\% of the power-law component to the total flux based on $Chandra$ observations of the source in 10 February and 5 June 2009 and in 15 April 2010 (note the proximity in time of the $Chandra$ observations to the ones analysed in this work). The sensitivity to detect and accurately model the power-law component is crucial to constrain the behaviour of the thermal component. In this sense, the higher effective area of \xmm\ compared to $Chandra$ and the longer exposures make the observations presented in this work more suitable to determine the contribution of the power law to the total flux. However, the different power law fluxes could be explained if the source varied between the $Chandra$ and \xmm\ observations.

We observe a significant decrease in the optical magnitude of the system. The most significant drop of intensity occurs after Obs~1 and is very small between Obs~2 and 4. This supports further the results from X-ray spectral fitting in the sense that any residual accretion present in Obs~1 has disappeared in subsequent observations. \citet{0748:hynes09apj} observed \src\ between November 2008 and January 2009 using Andicam and the SMARTS 1.3 m telescope. They found an average magnitude of R = 22.4 and J = 21.3 for the optical counterpart and a periodicity consistent with the orbital period of the system, indicating that at the time of the observations emission from the accretion disc and/or X-ray heated inner face of the companion star dominate the optical emission.  We do not observe any significant modulation in Obs~2 and 4, indicating that the emission from the accretion disc has most likely disappeared and the inner face of the companion star has subsequently cooled down. However, given the large errors of the measurements in Obs~2 and 4, the existence of a small modulation due to a remaining temperature gradient between the illuminated and dark face of the companion cannot be ruled out.

We attempted to constrain the value of the interstellar absorption with the high-resolution RGS spectra. This is important especially in the case that spectral variability in the power-law component is detected. Coupled variations between the power-law index and the value of \nh\ were found for Aql~X-1 and Cen~X-4 \citep{aql:campana03apj,cenx4:campana04apj} which could be explained if the power-law arises as shocked emission from the pulsar wind and the infalling material \citep{campana98aar}. Unfortunately, the absorption edges are not strong for these observations due to the low interstellar absorption in the direction of the source, $<$10$^{21}$ cm$^{-2}$, and therefore the value of \nh\ determined from the RGS spectra alone depends on the continuum used. However, since we only detect contribution of a power-law component in the first of the four observations, the variations in the value of \nh\ obtained are non-critical for our analysis.
 
We do not find absorption lines in the high-resolution RGS spectra, in agreement with expectations. The detection of lines from the NS surface in the spectra of quiescent NSs would allow the determination of the gravitational redshift \citep{brown98apjl,rutledge02apj} and are therefore of great interest. However, given the high spin and inclination of \src, the probability of detecting such features is very low \citep{chang06apjl,0748:lin10apj}.

\subsection{Constraints from the cooling and heating curves}
\label{sec:cooling}

The gradual decrease in thermal flux and neutron star temperature can be interpreted as the NS crust cooling down in quiescence after it has been heated during its long accretion outburst.  We found an e-folding time for this cooling of 133\,$\pm$\,88~d, a normalisation of 17\,$\pm$\,6~eV and a constant level of 109\,$\pm$\,2~eV when fitting the temperature decay curve with an exponential law. The normalisation and the constant level are consistent with those inferred from a similar fit to $Chandra$ (13.4\,$\pm$\,0.2~eV and 107.9\,$\pm$\,0.2) and $Swift$ (17.2\,$\pm$\,1.8~eV, 106.2\,$\pm$\,2.5~eV) observations. In contrast, the e-folding time although consistent within the errors, is smaller than the one inferred from $Chandra$ (192\,$\pm$\,10~d) and $Swift$ (257\,$\pm$\,100~d) observations \citep{0748:degenaar10mnras}. When fitting the decay curve with a power-law, we find parameter values of A\,=\,141.0\,$\pm$\,8.4~eV and B\,=\,-0.04\,$\pm$\,0.01, consistent with the values inferred from a similar fit to $Chandra$  (A\,=\,134.4\,$\pm$\,1~eV, B\,=\,-0.03\,$\pm$\,0.01) and $Swift$ (A\,=\,144.7\,$\pm$\,3.8~eV, B\,=\,-0.05\,$\pm$\,0.01) observations \citep{0748:degenaar10mnras}.

The cooling curve of \src\ is much flatter compared to the currently cooling NSs \ks\ and \mxb. \ks\ shows cooling consistent with a power-law decay with index  -0.125\,$\pm$\,0.007 still 8~years after the end of outburst \citep{1731:cackett10apj}. \mxb\ has cooled instead following an exponential decay with e-folding time of 465\,$\pm$\,25~d, reaching a temperature of 54\,$\pm$\,2~eV, which has remained constant for $\sim$1000~days \citep{1659:cackett08apj}. In contrast, the cooling curve of the NS \xte\ can be fitted with an exponential law and shows an e-folding time of 120\,$^{+30}_{-20}$~d \citep{1701:fridriksson10apj}, very similar to the value found for \src\ in this work. 

The decay time provides a measure of the thermal relaxation time of the NS crust \citep[e.g.][]{rutledge02apj, brown09apj}. In particular, observing the drop in luminosity permits the measurement of the thermal timescale of the crust and the relative magnitude of the drop tells us about the presence or absence of enhanced neutrino emission from the core. \citet{rutledge02apj} simulated the thermal relaxation of the NS crust in the transient \ks\ considering different impurity fractions and thus conductivities of the crust and the possible presence of "enhanced" core neutrino mechanisms such as direct Urca or pion condensation. Comparing the shapes of the curves in their Fig.~3 with Fig.~\ref{fig:decay-curve}, we conclude that \src\ is consistent with having a NS crust with a high thermal conductivity and low-impurity material, in agreement with \citet{0748:degenaar10mnras} and consistent with the other three cooling NSs  \citep{1731:wijnands02apj, 1659:wijnands04apjl, brown09apj, 1701:fridriksson10apj}.  

From the cooling curve we infer that the NS crust is close to reach thermal equilibrium with the core. Therefore, we can compare our results with theoretical predictions of the quiescent thermal luminosities of accreting NS transients. \citet{yakovlev04aa} computed the quiescent bolometric thermal luminosity as a function of long-term time-averaged mass accretion rate for several models of accreting NSs warmed by deep crustal heating. We calculated the average accretion rate during the last  11.5~years of outburst monitored with the RXTE/ASM. Considering only daily detections above 3~$\sigma$, we obtained an average count rate of 1.20~s$^{-1}$. We found an average ASM count rate of 0.68~s$^{-1}$ during the \xmm\ monitoring of \src\ in 2003, for which a bolometric (0.1-100~keV) unabsorbed flux of 8.44\,$\times$\,10$^{-10}$ erg cm$^{-2}$ s$^{-1}$ was obtained \citep{ionabs:diaz06aa,0748:boirin07aa}. Using this flux as a conversion factor we obtain an average bolometric flux of 1.49\,$\times$\,10$^{-9}$ erg cm$^{-2}$ s$^{-1}$ during the second half of the outburst. Comparing the thermal bolometric luminosity of 5.6\,$\times$\,10$^{33}$ (d/7.1 kpc)$^2$ erg s$^{-1}$ in June 2010 and the average accretion rate during outburst with the {\it heating curves} calculated by \citet{yakovlev04aa} (see their Fig.~3), we conclude that if the crust of \src\ has reached or is almost reaching thermal equilibrium with the core, a low to medium-mass NS with a near-standard cooling scenario is more likely than a high-mass star with significantly enhanced cooling.

We note that the values of the thermal bolometric flux from which we calculate the thermal bolometric luminosity are similar to the values inferred from $Swift$ data but higher than values from $Chandra$ data, as previously remarked by \citet{0748:degenaar10mnras}. This probably points to an offset in the flux calibration among these observatories.

\subsection{Constraints from spectral fitting to the thermal emission}

We attempted to constrain the mass and radius of the NS from spectral fits to the observed thermal emission from the NS surface. \citet{0748:zhang10mnras} performed a similar analysis but only with the first of the observations presented here. They found that fitting the thermal emission with two different NS atmosphere models, {\tt nsgrav} and {\tt nsatmos}, yielded similar results. Therefore we only used the {\tt nsatmos} model in our analysis. By fitting the four observations simultaneously we were able to reduce significantly the contours that define the mass and radius of the NS with respect to the analysis by \citet{0748:zhang10mnras}. 

\citet{steiner10apj} determined an empirical dense matter EoS from a  heterogeneous dataset of six NSs. They found significant constraints on the mass-radius relation for NSs, and hence on the pressure-density relation of dense matter. For example, for NSs where the photospheric radius  equals the NS radius, they found radii of 11.0 and 10.6~km for NS masses of 1.5 and 1.8~$\Msun$, respectively. For NSs where they allowed a photospheric radius larger than the NS radius, they found radii of 11.8 and 11.6~km for a NS mass of 1.5 and 1.8~$\Msun$, respectively. 
We obtain a mass of 1.78\,$\Msun$ and a radius of 13.7~km for a distance of 7.1~kpc and of 1.53\,$\Msun$ and 11.8~km for the lower limit of the distance of 5.9~kpc. Interestingly, our solution for a distance of 5.9~kpc agrees very well with the EoS found by \citet{steiner10apj} and would imply a medium-mass NS, in agreement with the mass derived by comparison with the heating curves obtained by \citet{yakovlev04aa} (see Sect.~\ref{sec:cooling}). In contrast, comparing the contours in Fig.~\ref{fig:contours} with predictions for the mass to radius relation for representative EoSs \citep[e.g.][]{lattimer01apj}, the solution for a distance of 7.1~kpc rules out most of the EoSs derived for an interior of nucleons and hyperons.

There are several factors which can include uncertainties in the calculation of the distance for \src. Firstly, \src\ is a high-inclination, dipping source, and as such there may be a changing partial obscuration of the NS and its expanding photosphere by the disc during the radius expansion of the type I bursts \citep{dist:galloway08mnras}. The ratio of the peak burst flux to the flux at touch-down (at the moment when the photosphere 'touches down' on to the NS) is expected to increase with increasing radius of the photosphere, with respect to low-inclination sources, and thus the distance could be overestimated. Taking this into account, \citet{dist:galloway08mnras} estimated a value for the distance to \src\ of 7.1\,$\pm$\,1.2~kpc. However, the value of the distance could be smaller if the obscuration at the touch-down flux were still underestimated. We note that in an analysis of all the dipping sources observed by \xmm\ \citep{ionabs:diaz06aa}, \src\ showed the least ionised atmosphere and a relatively cool plasma was present at all phases, indicating that obscuration in this source was larger than in other dippers.  This result was confirmed by an analysis of 600~ks of high-resolution RGS spectra of \src \citep{0748:vanpeet09aa}. Secondly, the distance calculations from type I X-ray bursts assume that at touch-down, the photosphere radius is equal to the NS radius. However, \citet{steiner10apj} argue that a photosphere with a radius larger than the NS provides internal consistency in their analysis and thus the assumption that the photospheric radius is equal to the NS radius is suspect. Further uncertainties arise from variations in the composition of the photosphere, the NS mass or variations in the typical maximum radius reached during the type I burst episodes (which affects the gravitational redshift, and hence the observed Eddington luminosity) \citep{dist:galloway08mnras}.

Therefore, we conclude that major uncertainty in the current limits on the mass and radius of \src\ are driven by the uncertainty in the distance estimate. Moreover, in order to obtain consistency between the mass obtained from the heating curves and the spectral fits to the thermal component, the distance to \src\ should be $\approxlt$6~kpc. Alternatively, a low/medium mass NS, albeit with a radius of 14--15~km, is possible at $\sim$7~kpc for harder EoS \citep[e.g.][]{lattimer07pr}. 

\subsection{A revision on the distance estimate to \src}
\label{sec:distance}

All current distance estimates to \src\ are based in the analysis of thermonuclear (type-I) X-ray bursts \citep[][]{dist:jonker04mnras,0748:wolff05apj,0748:ozel06nat,dist:galloway08apjs,dist:galloway08mnras}. This method assumes that the peak flux for very bright bursts reaches the Eddington luminosity at the surface of the NS, at which point the outward radiation pressure equals or exceeds the gravitational force binding the outer layers of accreted material to the star. Once the Eddington flux is reached, the spectral evolution during the first seconds shows a maximum in blackbody radius simultaneous with a minimum in color temperature, while the flux remains constant, indicating that the photosphere expands and the effective temperature decreases in order to maintain the luminosity at the Eddington limit. A large uncertainty in the theoretical Eddington luminosity arises from variations in the photospheric composition. A second large uncertainty arises from the fact that different bursts exhibit different peak fluxes. These differences may be caused by variable obscuration \citep[e.g.][]{dist:galloway03apj}. If this is true, the uncertainty will be significantly larger  for high-inclination sources, such as \src, for which strong absorption along the line of sight is common. Indeed, \citet{0748:asai06pasj} found that the burst profiles of \src\ in the soft energy band are highly affected by the presence of a photo-ionised plasma, whose ionisation degree changes largely by the strong X-ray irradiation of the burst. 
Since absorption along the line of sight reduces the observed flux, bursts with higher flux will be more reliable as distance estimators since they are less affected by obscuration. Further uncertainties include the mass of the NS and the gravitational redshift.

Up to now, three photospheric radius expansion bursts have been reported from RXTE on May 2004 and June and August 2005 \citep{0748:wolff05apj, dist:galloway08apjs}. The burst on May 2004 shows the largest peak flux. Based on spectral analysis of this burst, \citet{0748:wolff05apj} estimated a distance of 7.7~kpc (5.9~kpc) for a helium(hydrogen)-dominated burst photosphere. \citet{dist:galloway08apjs} used the three bursts to infer a distance of 8.3/(1+X) kpc, where X is 0 for solar abundances and ~0.7 for hydrogen-rich material. Further, they constrained the distance to 7.4 kpc arguing that the bursts short rise times and low duration suggest ignition in a He-rich environment. We note that the \citet{dist:galloway08apjs} obtain a peak flux for the brightest burst of 4.7\,$\times$\,10$^{-8}$  ergs~cm$^{-2}$~s$^{-1}$, 9\% lower than the flux obtained by \citet{0748:wolff05apj} due to the lower effective area of LHEASOFT v5.3, used in the latter paper \citep{dist:galloway08apjs}. 

We estimate that  Compton scattering in a highly ionised plasma (log~$\xi\sim$\,4) for a high-inclination source such as \src\ can reduce the flux by 1 to 10\% for a column density between 10$^{22}$ and 10$^{23}$ cm$^{-2}$. If the plasma has a lower degree of ionisation, the flux is reduced by a larger factor.
Therefore, if the flux of the burst on May 2004 is underestimated by 1--10\%, this translates into a distance of 6.8--6.5~kpc for a helium-dominated burst photosphere and of 5.2--5.0~kpc for a hydrogen-dominated burst photosphere. These values are consistent with an upper limit to the distance of $\approxlt$~7~kpc, required by physical EoSs (see Fig.~\ref{fig:contours}).
 
\section{Conclusions}
\label{sec:conclusions}

We analysed four observations of \src\ spanning 19 months after a significant decrease of accretion on the NS was detected. We conclude that:

- the emission above 3~keV decreased significantly within six months of the end of the accretion phase

- the NS crust has a high thermal conductivity and low impurity material

- the NS crust is very likely close to reach thermal equilibrium with the core

- the NS has a low-medium mass and is cooling by standard mechanisms

- a consistent solution between spectral fitting to the thermal spectra and the heating curves requires a distance of $\approxlt$6~kpc to \src.


\begin{acknowledgements}
Based on observations obtained with XMM-Newton, an ESA science
mission with instruments and contributions directly funded by ESA
member states and the USA (NASA). M. D{\'i}az Trigo thanks useful 
discussions with J. Tr{\"u}mper, R. Rutledge and F. Haberl.
\end{acknowledgements}


\bibliographystyle{aa}
\bibliography{mybib}

\end{document}